# 5G New Radio for Public Safety Mission Critical Communications

Jingya Li, Keerthi Kumar Nagalapur, Erik Stare, Satyam Dwivedi, Shehzad Ali Ashraf, Per-Erik Eriksson, Ulrika Engström, Woong-Hee Lee, Thorsten Lohmar

Ericsson

Contact: {jingya.li, keerthi.kumar.nagalapur, erik.stare, satyam.dwivedi, shehzad.ali.ashraf, per-erik.s.eriksson, ulrika.engstrom, woonghee.lee, thorsten.lohmar}@ericsson.com

*Abstract—* **Driven by increasing demands on connectivity to improve safety, situational awareness and operational effectiveness for first responders, more and more public safety agencies are realizing the need of modernization of their existing non-3GPP networks. 3GPP based cellular networks offer the unique opportunity of providing fast, reliable, and prioritized communications for first responders in a shared network. In this article, we give an overview of service requirements of public safety mission critical communications. We identify key technical challenges and explain how 5G NR features are being evolved to meet the emerging safety critical requirements, including enabling connectivity everywhere, supporting efficient group communications, prioritizing mission critical traffic, and providing accurate positioning for first responders.**

## I. Introduction

Fast, reliable, and secured connectivity is the key for public safety mission critical (MC) operations, where the availability of communication can be a question of life or death. Many of the existing public safety MC communications are based on narrowband land mobile radio (LMR) systems, which can only provide voice-centric services with very limited data capabilities. In addition, it is difficult to coordinate inter-agency operations since different public safety institutions typically use different frequencies and technologies.

To improve the situational awareness and operational effectiveness of first responders, there is a constant rise in demand on modernization of their existing non-3GPP networks to support the emerging advanced MC applications such as real-time MC video, remote control of drones/robots, and augmented reality with haptic feedback [1]-[3]. More and more telecom operators are requested by governments to provide MC services and coverage with their cellular mobile networks for first responders. Examples include the US national-wide public safety networks provided by FirstNet in partnership with AT&T, the Emergency Service Network critical communications system provided by Home Office in partnership with BT/EE in UK, and more recently, the 3GPP-based national MC mobile network that will be provided by Erillisverkot in Finland.

3GPP has developed and evolved the capability of 4G/5G systems for supporting public safety MC communications, ranging from application layer design, security solutions, system architecture design, to radio access network (RAN) design. The adoption of 3GPP standardized technology can facilitate MC communications among different safety organizations through a single, high-speed, and reliable wireless broadband network.

A survey of 4G long term evolution (LTE) based public safety networks is provided in [4], where spectrum aspects and LTE standardized solutions tailored to support public safety use cases are discussed. In [5], a comprehensive review of public safety services, virtualized and cloud-native 5G technologies, architectures and deployment options are presented. The focus of this article is on the design aspects of 3GPP 5G wireless access technology, known as new radio (NR), that support public safety MC communications. NR is a flexible radio interface that can meet a wide range of service requirements and deployments. An overview of fundamental NR concepts is provided in [6]. In this article, we describe the public safety MC service requirements, identify the key challenges to fulfil these requirements, and explain how the evolved NR features can be used to overcome these challenges. Especially, we take a deep dive into the latest NR features that play a crucial role in public safety communications.

## II. Use Cases and Challenges of Public Safety MC Communications

### A. Emerging Use Cases and Service Requirements

Public safety MC communication is shifting from voice-only services to more data driven internet of things (IoT) type of services. Some exemplary use cases include remote control of drones and robots, instead of using people, to gain real-time situational awareness or to deliver medical supplies; using haptic sensors in firefighters' personal protective equipment for quicker and safer search-and-rescue in a burning building with heavy smoke; and connected ambulance with support from a remote medical expert to save lives.

MC operations can be required in different network coverage situations, e.g., within terrestrial cellular network coverage, in partial terrestrial cellular network coverage, and out of terrestrial network coverage. And they may have different communication range requirements, e.g., localized communication, wide-area communication, and air-to-ground communications.

Table 1 QoS Characteristics of 3GPP-based MC Services

| Service | | 5QI values | Resource type | Priority level (lower value means higher priority) | Packet delay budget (including core network delay) | RAN delay budget | Packet error rate |
|---|---|---|---|---|---|---|---|
| MCPTT | Voice user plane | 65 | Guaranteed flow bit rate (GBR) | 7 | 75 ms | 65 ms | $10^{-2}$ |
| | Signaling | 69 | Non-GBR | 5 | 60 ms | 50 ms | $10^{-6}$ |
| MCVideo user plane | | 67 | GBR | 15 | 100 ms | 100 ms | $10^{-3}$ |
| MCData | | 70 | Non-GBR | 55 | 200 ms | 200 ms | $10^{-6}$ |

In 3GPP, MC services are divided into three categories, i.e., MC push-to-talk (MCPTT) [7], MCData [8], and MCVideo [9]. MCPTT is a voice-type service, whose data rate can range from 20 kbps to 70 kbps depending on the codec and PTT implementation. MCData includes services like short data service, file distribution, and data streaming for remote control of robots and drones. The data rate for MCData can range from 10 kbps to 1Mbps. MCVideo includes three different video modes: urgent real time video, non-urgent real time video, and non-real time video. Depending on the required video quality, the data rate requirement for MCVideo services can vary between 150 kbps and 5 Mbps.

The quality of service (QoS) characteristics for MCPTT, MCData and MCVideo are defined by 3GPP as the standardized 5G QoS Identifiers (5QI) [10], and they are summarized in Table 1. A 5QI value maps to a set of QoS characteristics, including resource type, default priority level, packet delay budget and packet error rate requirement. These characteristics are used as guidelines for an end-to-end QoS flow traffic forwarding treatment. The resource type determines if special treatment is required for a QoS flow to guarantee its performance requirement. The priority level indicates a priority of a QoS flow, which can be used as an input for admission control and scheduling of resources. The packet delay budget defines the allowed latency, for a packet delivery, between the user equipment (UE) and the core network. The packet error rate defines an upper bound for non-congestion related packet loss rate, which gives input to radio access link layer protocol configurations.

*B. Challenges of Public Safety MC Communications*

The capability of 5G NR is being enhanced in 3GPP to address the following key challenges of public safety use cases.

**Limitless Connectivity**: MC communications require coverage/capacity that is beyond what is typically needed for commercial services. If required, the MC services must be delivered in time regardless of whether the UE is within the terrestrial network coverage or not [2], [7]-[9]. In disaster scenarios or underground situations, the cellular coverage/capacity provided by the existing network at a particular geographical location may not be enough, or the terrestrial network may not be available. In such situations, additional means are required to ensure real-time and reliable MC communications.

**Efficient Group Communications**: First responders typically work in groups and require group communications to efficiently coordinate their operations [7][6]-[9]. In large public safety missions, where the system can be highly loaded by large number of first responders and commercial mobile users, it is important to optimize the system to deliver the MC services for each group with the required quality, so that more MC UE groups and commercial users can be served simultaneously.

**Prioritization of MC Traffic**: Due to increased spectrum needs and seeking of cost-efficient MC solutions, more and more public safety agencies are considering the use of prioritized access to mobile network operators' spectrum. When using a shared 5G network for MC communications, it is crucial to ensure the network accessibility and service quality in high load situations [2]. The network must be able to identify early and prioritize the access requests from MC UEs and maintain the service quality during network congestion. There might be situations where all commercial users are barred from accessing the network, but the system capacity is still not enough to serve all requested MC communications; or situations where the first responders are connected to a deployable network (e.g., cells-on-wheels/wings), which has much lower capabilities than traditional cellular networks. In such cases, the access control mechanism needs to wisely determine which of the prioritized MC users and services shall access the system. For this, priority differentiation between different MC users and services are needed.

**Accurate Positioning for First Responders**: Real-time accurate positioning of first responders is essential for improving their safety and situational awareness. MC situations require positioning solutions that can provide on-demand accurate location of first responders in all situations, regardless of the availability of satellite or cellular network coverage [2], [3]. The most challenging positioning scenario for public safety is 3D indoor positioning of firefighters in a burning residential building with broken power supply, where the indoor base stations (BSs) or nodes cannot work due to lack of power supply, and the global navigation satellite system (GNSS) signals are too weak to provide indoor positioning with sufficient accuracy. The level of positioning accuracy required in this case is around 1 meter in horizontal and 2–3 meters in vertical directions (to identify the floor level), which is more stringent than the regulatory requirements for commercial users, but less stringent than that required by industry IoT use cases.

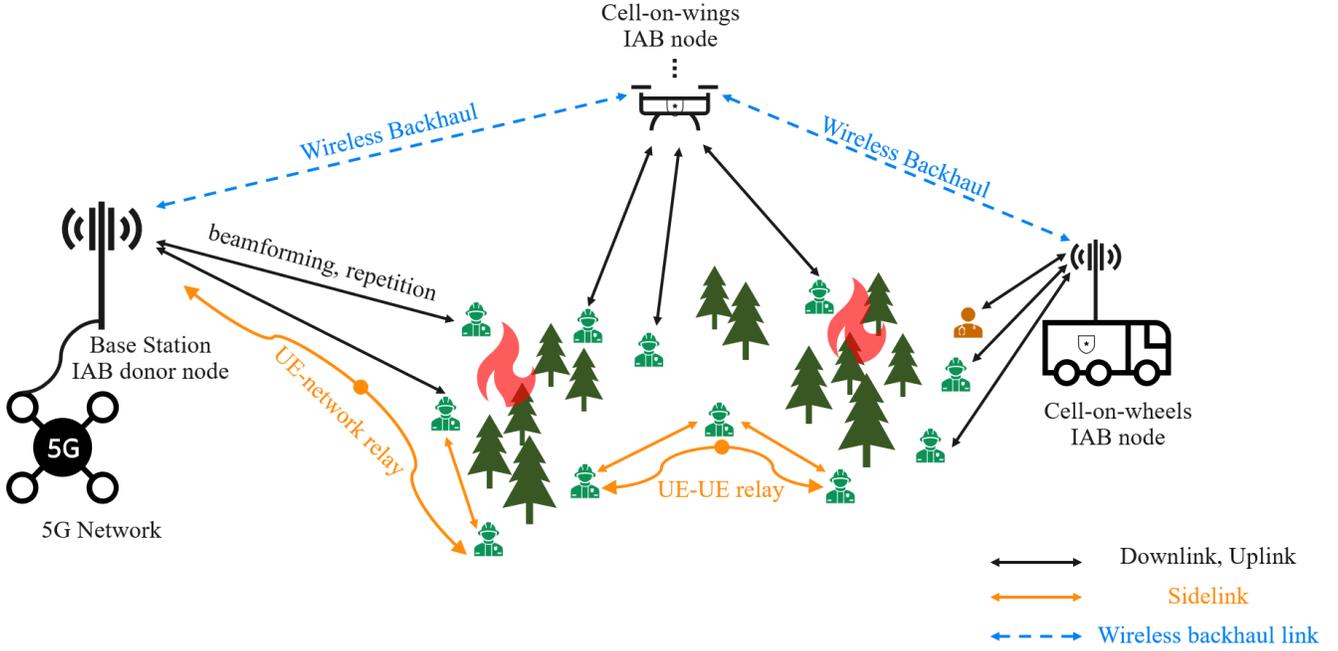

Figure 1 5G NR coverage solutions to provide limitless connectivity for public safety MC communications.

## III. 5G NR COVERAGE SOLUTIONS FOR PROVIDING LIMITLESS CONNECTIVITY

5G NR has developed rich features for network coverage extension, which can be applied for providing good coverage for MC communications. As shown in Figure 1, examples of NR coverage solutions include using *beamforming* to increase the signal strength for intended UEs in a specific area, applying *repetition* to transmit the same data multiple times, configuring an UE with *high-power classes* so that it can transmit signals with a higher power, and using *integrated access backhaul* (IAB) for multi-hop network relaying or for enabling deployable networks. In addition, NR supports *sidelink* (SL) including SL relaying to provide coverage in areas where network infrastructure is not available or damaged. In the following, we explain in detail on how the IAB and SL features designed in 5G NR can be used to provide connectivity for MC UEs when they are out of cellular network coverage.

### A. IAB-based Deployable Systems for On-Demand Coverage

Drones carrying BSs, often referred to as cells-on-wings in public safety community, offer great flexibility for establishing temporary networks [11]. They can be used either separately or in combination with the land-based portable BSs, e.g., cells-on-wheels, for providing temporary coverage to first responders in situations where the network infrastructure is damaged or not available. However, in practice, there are two major challenges that need to be addressed, i.e., the limited on-board power of drones, and the backhaul connectivity between the cells-on-wings and the core network.

Compared to the satellite-based backhaul solution, which is costly and doesn't scale well to high-capacity and time-sensitive traffic demands, the NR IAB feature enables flexible deployment of cells-on-wings due to its combined wireless backhauling and access capabilities [12]. In an IAB deployment, an IAB child node providing wireless access to the users is connected, through a wireless backhaul link realized using the cellular (Uu) interface, to an IAB donor node having core connectivity. The standardized solution for IAB is based on the central unit (CU)/distributed unit (DU) split architecture of a 5G BS (i.e., gNB). The CU terminates the less time-critical packet data convergence protocol (PDCP) and the radio resource control (RRC) protocol. The DU terminates the latency-sensitive lower layer protocols, i.e., radio link control (RLC), medium access control (MAC) and the physical layer. The IAB donor node hosts the CU and a donor DU. The donor DU provides a backhaul link to an IAB child node, and it also has the capability to provide NR access to ordinary UEs. The IAB child nodes have a mobile termination (MT) that establishes connectivity with a donor DU to provide backhaul connectivity as shown in Figure 2 (a). The IAB child node hosts a DU that is controlled by the CU on the IAB donor node; the F1 interface connects the CU of an IAB donor node to the DUs of its IAB child nodes. In the following, we present how the NR IAB feature can be used to enable flexible deployment of cells-on-wings.

#### 1) Portable Network Node Integration

An IAB integration procedure is performed to integrate an IAB child node with the network [12]. To begin with, the MT of the IAB child node makes a usual cell search using procedures like a regular UE. It connects and camps on a cell established by the DU of the IAB donor node. The MT makes a normal RRC-connection with the CU of the IAB donor node, where authorization/authentication is also executed. An F1 connection is established between the DU of the IAB child node and the CU of the IAB donor node. Using the F1 connection, the DU of the child node is configured with necessary parameters to provide wireless access to UEs or to provide backhaul for the next IAB node in a multi hop deployment.

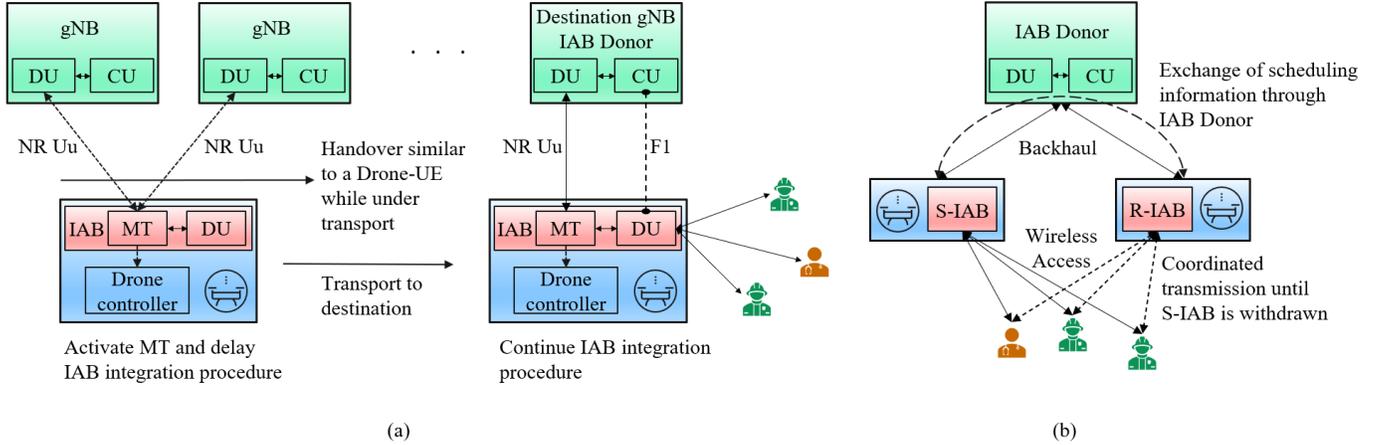

Figure 2: a) Integration of a Drone IAB BS. b) Seamless IAB node replacement.

When a drone is equipped with an IAB child node to provide on-demand wireless access to MC users, it can introduce interference during its flight to the destination if the integration procedure is done before the flight. The architecture split of the IAB solution and the integration procedure can be exploited to overcome this challenge by delaying the F1 connection establishment while the drone is moving, as shown in Figure 2 (a). The integration procedure after the MT has connected to a gNB can be put on hold until the drone carrying the IAB child node has reached its destination. The integration procedure is resumed when the drone has reached its destination. Furthermore, the MT can be used as an interface between the network and the drone for remote control of the drone, which eliminates the need for a separate UE to control the drone.

*2) Portable Network Node Replacement*

In MC situations, a portable IAB child node may have to be replaced by another. For instance, it would be beneficial to replace a current serving IAB cell-on-wings (S-IAB) that is about to run out of power by a replacement IAB cell-on-wings (R-IAB) without service interruption. The UEs being served can be handed over to an R-IAB using the handover procedures adapted for IAB. Further enhancements can be employed to perform a seamless handover of MC users between the IAB child nodes sharing the same IAB donor to prevent potential service interruptions caused by handovers. Such a replacement can be performed by using coordinated transmission and reception from both the S-IAB and R-IAB before withdrawing the S-IAB, as illustrated in Figure 2 (b). The duplication of data necessary for coordinated transmission and the exchange of scheduling information can be achieved through the common IAB parent node that serves both child IAB nodes.

*B. Sidelink for Out-of-Coverage Communications*

5G NR SL interface to support direct communication between UEs is designed in 3GPP to complement the Uu interface, especially in areas without the network coverage. 5G NR SL supports many advanced features which are necessary for MC communications [13]. Such features include the support of unicast and groupcast communications. Unicast communication provides better QoS control and interference management for MC services. Furthermore, the groupcast and unicast communications also allow hybrid automatic repeat request (HARQ) based retransmissions to improve reliability and spectral efficiency. Sensing based autonomous resource allocation by the UE is enhanced to consider aperiodic traffic type, which is particularly relevant for latency critical MC communications. Given the critical nature of MC traffic and different service priorities, pre-emption as introduced in 5G NR SL is a beneficial feature for MC services. Additionally, 3GPP is currently working on power efficient SL operation which is important for first responder devices. SL features enabling power efficient operation include the introduction of discontinuous reception and partial sensing operation. Also, SL based relaying is being studied to improve the network coverage and availability.

## IV. 5G NR MULTICAST FOR EFFICIENT GROUP COMMUNICATIONS

With NR Release 17, 5G will support *group communication* to efficiently deliver multicast and broadcast services (MBS), via IP multicast data, to a group of users. With 5G multicast, the data is transmitted in relevant cells to UEs that have joined the related MBS session. The IP multicast data of an MBS session is associated with one or more QoS levels mapped to the agreed QoS requirements, which the RAN needs to fulfill.

For 5G multicast, the RAN uses one or more multicast radio bearers (MRB). An MRB fills the same role as a traditional data radio bearer (DRB) for unicast and delivers IP multicast data to UEs. The MRB is split into a point-to-multipoint (PTM) and a point-to-point (PTP) "leg". The PTM leg supports true PTM transmission. The PTP leg behaves essentially as traditional bi-directional unicast but is logically separated from the unicast functionality of the DRB, which may be used in parallel. Both legs, PTM and PTP, inherit most parts of the legacy NR protocol stack, with PDCP, RLC, MAC and PHY.

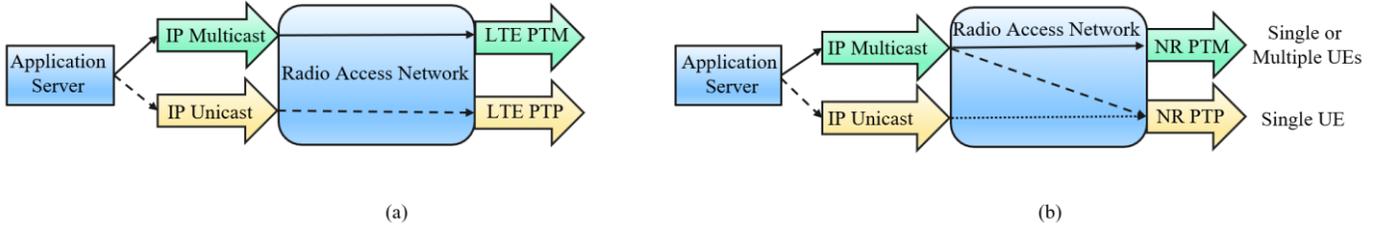

Figure 3: a) LTE group communication solution: Switch between IP unicast (LTE PTP) and IP multicast (LTE broadcast/PTM) controlled by AS in LTE. b) NR group communication solution: Switch between PTM and PTP controlled by RAN in NR.

Figure 3 illustrates one important difference between LTE broadcast and NR multicast solutions. In LTE the broadcast transmission in a service area is set up in a semi-static way with the application server (AS) making decisions on whether to transmit via IP unicast (LTE PTP) or via IP multicast (LTE broadcast). In NR, IP multicast is almost always used for the MC data and how to deliver this IP multicast to the end UEs is instead decided by the RAN independently (e.g., PTM/PTP decision). However, also with NR, the AS can exceptionally and outside the 5G multicast-broadcast services (5MBS) system, decide to use IP unicast/PTP for the MC traffic. In contrast to the relatively inefficient semi-static set up in LTE, the NR design allows for a more adaptive group communication solution controlled by the RAN. For example, the RAN can dynamically select different solutions (e.g., NR PTM transmission, NR PTP transmission, multi-user MIMO, link adaptation, beamforming, retransmission, repetition, etc.) based on the current network situation (e.g., number of UEs, UE locations, and channel condition, etc.). Therefore, compared to LTE, the NR multicast design can improve the radio resource management, providing a more efficient trade-off between QoS/coverage and spectral efficiency. It can also improve service continuity during handover or switching between PTP and PTM.

To minimize the feature implementation cost at the user equipment, the design of 5G NR multicast features are largely built on the already feature-rich NR unicast features. Examples include using the same numerology, the same physical channel structure, etc. Channel State Information (CSI) reporting from the UE will also be directly reused. With CSI report, the network may adapt the transmissions to the actual link quality and may then take all UEs in a PTM group into account to optimize the transmission. To improve the reliability of MBS delivery, in 5G NR, both NR PTM and PTP legs support HARQ retransmissions based on UE feedback. Within the PTM leg, retransmissions can use either PTM or PTP, whichever is the most efficient.

For handling mobility of MBS users in RRC Connected state, methods like the ones used for unicast users can be used. If the same multicast session has not started in the target cell, but the target cell supports 5G MBS, the UE context will be transferred to the target cell, which connects with the multicast session, so that the target cell is prepared for the UE handover. If the same multicast session is already going on in the target cell, then, there is no need for the target cell to connect with the multicast session - the UE context is transferred to the target cell and possibly lost PDCP packets during handover can be retransmitted to avoid data loss.

## V. 5G ACCESS CONTROL AND ADMISSION CONTROL FOR PRIORITIZING MC TRAFFIC

Prioritization of MC traffic can be achieved by using the 5G access control framework, which is divided into four main functionalities, namely, cell barring and reservation, unified access control (UAC), random access control, and admission control. In case of resource limitation, these functionalities can be used to selectively limit the incoming traffic to the system at different connection setup stages. An overview of 3GPP based 5G system accessibility differentiation and control is provided in [14]. Here, we present the flow of 5G access control mechanisms that can be used for MC services (see Figure 4).

UAC is an access barring mechanism to block access requests originated at the UE side. Before a UE sends an access request to a network, the UE shall check the broadcasted UAC related barring information received from the gNB and determine if its access attempt is allowed. In NR, each access request is associated with one *access category* (service type) and one or more *access identities* (user profile). A first responder can be configured as an MC UE with a standardized *access identity*. Utilizing this parameter, in a high traffic load situation, the network can configure the UAC barring information so that the connection requests from MC UEs are allowed, but the requests from commercial users are barred with a certain probability. NR also supports operator-defined access categories, which are configured with criteria according to the operator needs for service differentiation. For example, an operator-defined access category can be configured to be associated with a network slice used for MC services. Thereby, the UAC barring information can be configured based on the congestion level of this network slice [14].

If an access request from a user is not barred, then the user can initiate a contention-based random-access (CBRA) procedure to establish a connection to the network, as shown in Figure 4. The *establishmentCause* carried on Msg3 indicates the reason for the access request. To support identification of MC UEs in initial access, a new establishmentCause, *mcs-PriorityAccess,* has been introduced in NR. Therefore, the gNB can identify an MC UE based on the establishmentCause received during CBRA procedure. Then, it can accept the access requests from MC UEs and reject the requests from commercial UEs in case of resource limitation. To achieve lower latency and higher probability of success with the random-access procedure, NR also supports the prioritized first message transmission with a larger power ramping step and shorter back off time for MC UEs. However, this prioritized first message transmission feature cannot be used for assisting the network to early identify an MC UE at the first step of a CBRA procedure in initial access. To achieve this, CBRA enhancement to support different Msg1 configurations between MC UEs and commercial UEs can be considered in future NR releases.

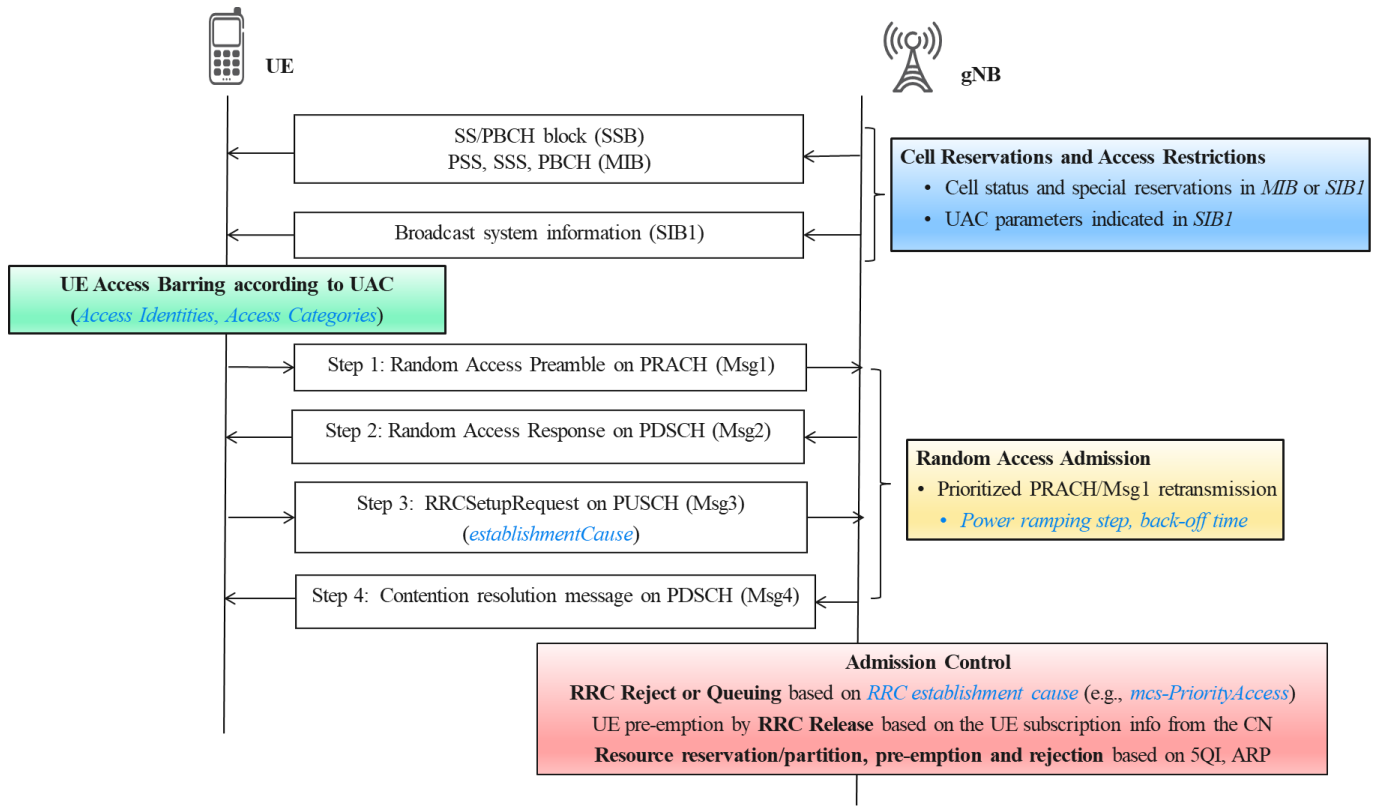

Figure 4 Flow of 5G access control and random access for prioritizing mission critical traffic.

Once a UE is connected to a 5G network, the network can acquire more QoS and priority related parameters of a UE and its required service. These parameters include, the network slice ID if it is preconfigured for the UE; the 5QI value; and the allocation and retention priority (ARP) that contains information about the priority level, the pre-emption capability and the pre-emption vulnerabilities. Based on these priority identifiers, the RAN scheduler can perform admission control mechanisms like differentiated traffic management and load balancing to further prioritize important MC traffic flows and pre-empt non-MC traffic when needed.

## VI. 5G NR POSITIONING FOR FIRST RESPONDERS

Positioning plays a vital role in improving the safety and situational awareness of first responders. Compared to LTE, in 5G NR, positioning signals and measurements have been enhanced to provide better positioning accuracy for both the indoor and outdoor scenarios [15].

5G NR supports two new reference signals for positioning, including downlink positioning reference signal (DL-PRS) and uplink sounding reference signal (UL-SRS). Compared to LTE, DL-PRS has a more regular structure, and it can be configured with a larger bandwidth, which allows better accuracy for timing-based positioning. 5G NR has also introduced the support of new positioning methods, including multi-cell round-trip time (multi-RTT) based positioning and angle-based positioning. The multi-RTT based positioning method is robust against network time synchronization errors since it utilizes both UL-SRS and DL-PRS measurements to estimate the location of a UE, unlike the popular time difference of arrival (TDOA) based methods where time synchronization among BSs is assumed. The angle-based positioning methods utilize the angle information, i.e., downlink angle of departure (DL-AoD) or uplink angle of arrival (UL-AoS), together with the network deployment and configuration information to estimate the location of a UE. Leveraging the benefits of multi-antenna beamforming features at mmWave, NR DL-PRS or UL-SRS for positioning can be transmitted in beams. Beam based DL-PRS or UL-SRS for positioning transmission enables the DL-AoD or UL-AoA estimation, and it also improves the coverage of the positioning reference signals.

As discussed in Section II, one challenging scenario for positioning of first responders is 3D location of firefighters in a burning building with broken power supply. One possible solution for this scenario is to use deployable BSs to provide additional communication links and positioning infrastructure to firefighters on site, and thereby, increasing the positioning accuracy provided by the cellular network. This can also provide benefit of locating any person carrying an NR device that are trapped in the building in unknown location.

Consider a positioning situation shown in Figure 5, where first responders arrive in a van equipped with equipment to establish communication with first responders when they get into the building on fire. This van in the figure is called command and control center (CCS). Multiple drone-BSs can act as positioning reference nodes (PRNs) and assist in keeping track of the responders' location. Thanks to the flexibility of drones, these deployable PRNs can be placed at different altitudes, which is especially beneficial for improving the positioning accuracy in the vertical domain. The location of drone-BSs can be estimated by using existing positioning methods. Cameras and sensors equipped on drone-BSs can also be used to provide additional data for positioning of first responders. Communications between different drone-BSs can be achieved by using the IAB feature as described in Section III. The first responders are fitted with a NR transmitter, which periodically transmits UL-SRS for positioning specified in Release 16. The drones communicate among themselves and are time synchronized. These drones do receive the UL-SRS for positioning and collect reference signal time difference measurements. The location server can be inside the CCS and computes the location of the first responders and thus helps coordinating the rescue operation.

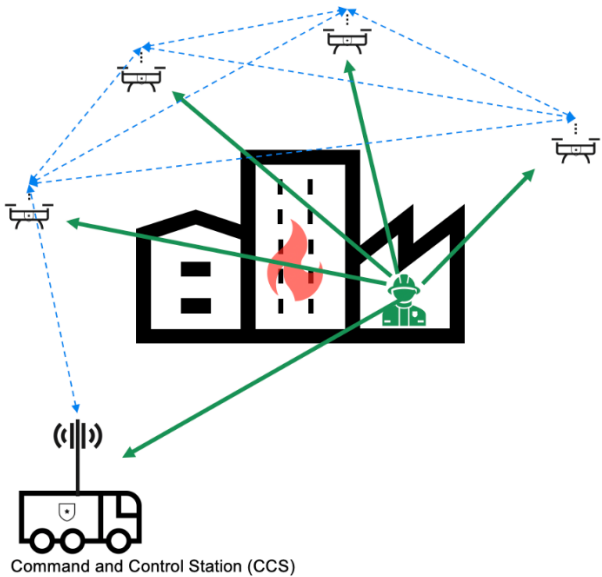

Figure 5 Illustration of a possible positioning setup for estimating positions of first responders inside the building on fire. The drone-BSs are time synchronized and receive UL-SRS from the rescue personnel. UL-TDOA positioning can be used to locate first responders.

Considering the drone's great flexibility, i.e., a system where measurement data can be dynamically collected, the NR positioning accuracy using deployable drone-BSs can be further improved by well-designed artificial intelligence-based approaches. The underlying philosophy for these approaches is twofold; first, drones learn the environment and apply statistical inference to the measurements for positioning based on data cleansing methods, e.g., denoising autoencoder; second, drones adaptively move themselves to acquire better measurements for positioning based on strategy selection methods e.g., deep reinforcement learning.

## VII. CONCLUSIONS AND FUTURE WORK

In this article, we provide an overview of the 3GPP NR features in supporting public safety MC communications. Compared to 4G LTE, 5G NR has introduced a set of *new features* for supporting advanced public safety use cases, including

- Enhanced coverage solutions for providing limitless connectivity, e.g., multi-hop IAB for flexible network coverage extension or portable IAB for providing temporary connectivity; Unicast and groupcast SL operations for enhanced reliability and improved energy efficiency.
- RAN controlled multicast to enable dynamic switching of different group communication solutions according to the current network load/resource situation.
- New service/subscriber identifiers and access control features for prioritizing network accessibility of MC users and maintaining the service quality of critical traffic.
- New positioning reference signals and methods to provide better positioning accuracy for first responders.

Although 5G NR offers many powerful features that can secure prioritized and reliable MC communications, it will take time before the ecosystem is mature to make the complete transition from LMR to 5G NR for public safety. Thus, it is important to study solutions for improved interoperability between LMR and NR based public safety networks. Meanwhile, the NR standard will continue to evolve in coming releases for enhancing its support for public safety. Some possible directions include access control and random-access enhancements to support early identification and finer differentiation between MC user/services, enhancements to support efficient group communications in a large area, and better support of using drones for improved situational awareness.

8[8] 3GPP TS 22.282 V16.4.0, "Mission Critical Data (MCData); Stage 1," Dec. 2018.
[9] 3GPP TS 22.281 V16.0.0, "Mission Critical Video (MCVideo); Stage 1," Oct. 2018.
[10] 3GPP TS 23.501 V17.0.0, "System Architecture for the 5G System (5GS); Stage 2," Mar. 2021.
[11] K. P. Morison and J. Calahorrano, "FirstNet Case Study: How FirstNet Deployables are Supporting Public Safety," Police Executive Research Forum, Oct. 2020. Available at: https://www.policeforum.org/assets/FirstNetDeployables.pdf
[12] C. Madapatha, B. Makki, C. Fang, *et al*., "On Integrated Access and Backhaul Networks: Current Status and Potentials," in *IEEE Open Journal of the Communications Society,* vol. 1, pp. 1374–1389, Sep. 2020.
[13] S. A. Ashraf, R. Blasco, H. Do, *et al*., "Supporting Vehicle-to-Everything Services by 5G New Radio Release-16 Systems," in *IEEE Communications Standards Magazine*, vol. 4, no. 1, pp. 26–32, Mar. 2020.
[14] J. Li, D. D. Penda, and H. Sahlin, *et al*., "An Overview of 5G System Accessibility Control and Differentiation", accepted by *IEEE Communications Standards Magazine*. Available at: https://arxiv.org/ftp/arxiv/papers/2012/2012.05520.pdf
[15] S. Dwivedi, R. Shreevastav, F. Munier, *et al*., "Positioning in 5G Networks", accepted by *IEEE Communications Magazine*. Available at: https://arxiv.org/pdf/2102.03361
BIOGRAPHIES

**Jingya Li** (jingya.li@ericsson.com) is a Master Researcher at Ericsson Research, Gothenburg, Sweden. She currently leads 5G/6G research and standardization in the areas of public safety mission critical communications. Jingya has led different back-office feature teams for 5G New Radio (NR) standardization. Her contribution to standardization has been on 5G NR for public safety, initial access, remote interference management and cross link interference handling, vehicle-to-anything communications, and latency reduction. She is the recipient of the IEEE 2015 ICC Best Paper Award and IEEE 2017 Sweden VT-COM-IT Joint Chapter Best Student Journal Paper Award. She holds a Ph.D. degree (2015) in Electrical Engineering from Chalmers University of Technology, Gothenburg, Sweden.

**Keerthi Kumar Nagalapur** (keerthi.kumar.nagalapur@ericsson.com) is a senior researcher at Ericsson Research, Gothenburg, Sweden. His current research activities focus on 5G NR for public safety communications and high-speed train communications. He holds a Ph.D degree in Electrical Engineering from Chalmers University of Technology, Gothenburg, Sweden.

**Erik Stare** (erik.stare@ericsson.com) holds an M.Sc. degree (1984) in telecommunication from KTH, Sweden. Between 1987 and 1992 (at Swedish Telecom) and 1992 and 2018 (at Teracom), he was involved in developing systems and standards for digital terrestrial broadcasting. Since 2018 he has held a Master Researcher position at Ericsson Research, working on 3GPP NR-based positioning and as team leader for 5G/NR multicast/broadcast standardization, including being a 3GPP RAN1 delegate.

**Satyam Dwivedi** (satyam.dwivedi@ericsson.com) is a master researcher at Ericsson research. His research areas are positioning and propagation. He was a researcher at KTH in Stockholm, Sweden. He holds Master and PhD from the Indian Institute of Science, Bangalore, India.

**Shehzad Ali Ashraf** (shehzad.ali.ashraf@ericsson.com) is with Ericsson Research, which he joined in 2013. He holds an M.Sc. in electrical engineering from RWTH Aachen University. Since joining Ericsson, he has been deeply involved in 3GPP standardization on 5G radio access technology. Currently, he is acting as a project manager for a research project looking into connectivity solutions for transportation industries and public safety.

**Per-Erik Eriksson** (per-erik.s.eriksson@ericsson.com) is a Master researcher at Ericsson Research. His current focus is research on fronthaul and backhaul solutions for 5G and future 6G. Per-Erik was also part of the research team that invented Ericsson Radio DOT. Since joining Ericsson in 1989 he has also been involved in DSL research and contributed actively in the standardization of ADSL2, VDSL2 and G.fast. Per-Erik holds an M.S. in electronic engineering from the Royal Institute of Technology, Stockholm.

**Ulrika Engström** (ulrika.engstrom@ericsson.com) joined Ericsson Research (ER), Gothenburg. Sweden, in 1999. She holds a Ph.D. (1999) and M.Sc. (1994) in Physics from Chalmers University of Technology, Gothenburg, Sweden. Ulrika has since then, during more than 20 years, driven key research projects and studies (both externally and internally) targeting 4G, 5G and new applications for mobile cellular systems towards 6G at ER. Her key research interest includes new use cases, with focus on deployment, spectrum, and sustainability aspects of mobile cellular systems. She is a Senior Researcher at ER.

**Woong-Hee Lee** (woonghee.lee@ericsson.com) is an Experienced Researcher with Ericsson Research, Stockholm, Sweden. He received the B.S. degree in electrical engineering from the Korea Advanced Institute of Science and Technology (KAIST), Daejeon, South Korea, in 2009, and the Ph.D. degree in electrical engineering from Seoul National University, Seoul, South Korea, in 2017. He was an Experienced Researcher with the Advanced Standard Research and Development Lab., LG Electronics CTO Division, Seoul. From 2019 to 2020, he was a Postdoctoral Researcher with the Department of Communication Systems, KTH Royal Institute of Technology, Stockholm, Sweden.

**Thorsten Lohmar** (thorsten.lohmar@ericsson.com) is graduated from the Technical University of Aachen (1997), where he also received his PhD (2011) in Electrical Engineering. Thorsten joined Ericsson in Germany in 1998 and was working for several years in different Business and Ericsson Research units. He worked on a variety of topics related to mobile communication systems and led research projects specifically in the multimedia technologies area. He is focusing on video delivery (downlink and uplink) and delivery optimizations, including broadcast distribution of media. He is currently working as Expert for Media Delivery.